\documentclass[pra,aps,showpacs,amsmath,amssymb,preprint,endfloats]{revtex4}
\usepackage{graphicx}
\usepackage{mathrsfs}
\usepackage{wasysym}
\usepackage{textcomp}
\usepackage{pifont}
\usepackage{stmaryrd}

\begin{document}

\title{Energy and information flow in superlensing}

\author{E. Fourkal}
\email{eugene.fourkal@fccc.edu}
\author{I. Velchev}
\affiliation{Department of Radiation Physics, Fox Chase Cancer Center,
Philadelphia, PA 19111, U.S.A.}
\author{A. Smolyakov}
\affiliation{Department of Physics and
Engineering Physics, University of Saskatchewan, Saskatoon, Canada}
\date{\today}

\begin {abstract}
In the superlens problem, the presence of the reflected wave plays a critical role  in the energy and information transfer from the object to recording device: both
incident and reflected evanescent waves are required to insure nonzero energy/information  flux from the object to the recorder. Therefore an optimization is required between the image quality (characterized by a transfer function) and information transfer rate (characterized by the energy/information flux). It is shown that the introduction of the recording device may lead to the deterioration in the image quality as compared to the ideal case when no recording device is present. The decline in the image quality is due to the generation of phase-shifted reflected evanescent wave between the lens and the recorder. It is also shown that the magnitude of the energy flux depends on the amount of dissipation in the lens and the recorder in a non-monotonous way.
\end {abstract}


\maketitle

\section{Introduction}

Propagation of the electromagnetic radiation in materials with negative dielectric permittivity (and permeability) or the so called left-handed materials (LHM) has attracted great deal of attention in recent years~\cite{kindel75,dragila85,bliokh05,fang05}. The increased interest in the properties of such media has been driven by their potential applications in various branches of science and technology, ranging from development of terahertz computer chips and next generation of integrated opto-electronic devices to ultra-sensitive molecular detectors (used in bio-molecular imaging applications), and invisibility cloak\cite{barnes03,maier05,smith04}. One additional application of these materials which has been widely studied in recent years is related to the possibility of creating the so called superlens: a sub-wavelength optical imaging system beyond the diffraction limit~\cite{fang05,pendry00}. 

The superlens phenomenon is essentially based on amplification of evanescent
components of the object spectrum  (these are the components that carry
sub-wavelength information about an object), facilitated by the excitation
of surface plasmons \cite{pendry00,gomez-santos03}. This part of the signal spectrum is
normally lost in the standard optical devices resulting in the diffraction
limit.  A simplified one dimensional object can be represented in the form 
$F(y)=\int F_{k}\exp \left( ik_{y}y\right) dk_{y}.$ The components with $%
k_{y}>\omega /c$ constitute the evanescent part of the spectrum and the components
with $k_{y}<\omega /c$ are propagating. In a standard optical device, only
the propagating components carry information from the object plane to the
image plane. The superlens on the other hand, has a property that both propagating and
evanescent parts of the spectrum can be transmitted via such a lens. The transmission of the evanescent part of the spectrum becomes possible via the amplification of the evanescent waves due
to the resonant excitation of the surface modes. 

The transfer of  information costs energy\cite{bekenstein81,bekenstein88}. 
As shown by Bekenstein\cite{bekenstein81}, there exists a universal bound on
the energy cost per bit of information transferred $E/I>\hbar \ln [2]/\pi
\tau $, which means that larger information transfer rates ultimately
require higher energy transmission (it also confirms the fact that it is
impossible to transfer information without spending some energy\cite%
{shannon48}). Therefore, this necessarily means that evanescent waves
carrying the information about sub-wavelength structure of an object have to
transfer energy. Depending on the "complexity" of an object (percentage
that evanescent components take in its Fourier spectrum), different amount
of energy flux is needed in order to transmit this sub-wavelength
information from an object plane to an image plane. There is an important
difference between the energy transport by the propagating ($k_{y}<\omega /c)
$ and evanescent $\left( k_{y}>\omega /c)\right)$ parts of the spectrum. For
propagating waves, the energy flux is finite and can be calculated
independently of the receiver. In other words, a finite energy flux can be
set up in a half infinite region so that an ideal recording device can be
imagined as that having a black body type absorption properties, receiving all of the available energy/information. In the evanescent part of the spectrum however, the situation is radically different. The energy flux in a half infinite region (where there is only one evanescent component $\exp \left(
-\alpha x\right) $ present) is identically zero. The second component $\exp
\left( \alpha x\right) $ with a proper phase shift is required to set a finite energy transport\cite{kolokolov92}. The existence  of the second component and the value of the phase shift are intrinsically related to the properties of the recording device. For example, in the absence of absorption in the recording device the energy flux in the evanescent part of the spectrum is identically zero. Thus, the recording device becomes an integral part of the imaging system in the process of capturing the energy and information, carried by the evanescent waves.     

Typically the quality of the lens is characterized by its transfer function (an ideal imaging system would have the transfer function equal to unity for all $k_y$, see below in the Section 2). The transfer function alone however ignores the issue of the transmitted energy flux (hence the amount
of transmitted information). This is not important for propagating waves as
the closeness of the transfer function to unity is the only criterion that determines the quality of the lens.
However, for the evanescent part of the spectrum, the ideal transfer function$(\tau =1$) does not guarantee sufficient energy/information flux. Therefore, for the subwavelength imaging the optimization of the energy flux
(determined by the dissipation in the recording device) becomes necessary. Discussion of these issues is a subject of the present paper.          

A generic imaging system consists of three main components: (1) the source of information carrying medium (electromagnetic waves, sound waves, etc.), (2) the lens, whose role is to apply proper phase adjustments to different spatial components of the spectrum and (3) the recording device, whose role is to capture the information, carried by mediating waves. By its purpose, the recording device is necessarily dissipative since it has to imprint the arriving information into "itself". The presence of the recording device also creates a reflected wave that ultimately ensures the energy and information flow between the lens and rd. 

In many previous research studies\cite{pendry00,fourkal07,podolskiy05} concerning image formation by the so called Pendry lens (a lens made out of a material having negative permittivity but positive permeability) the authors investigated lens's optical transfer function as well as its ability to focus an image in the absence of the recording device. This case however describes unrealistic situation, since therein obtained optical transfer function does not contain the influence of the recording device. As shown in Ref.~\cite{vinogradov05}, the introduction of the recording device may in some cases significantly deteriorate the quality of a resultant image (see below), which is due to the generation of a reflected wave by the lossy recorder. 

Here we investigate the role of the IRD (Image Recording Device) in the energy and information transfer between the object and image planes. We show that the level of image deterioration depends on an interplay between the amount of information needed to describe the imaged object (in a given time period) on one hand and the optical transfer function of the lens on the other. In particular, we study the case of a silver film (negative permittivity) deposited on another medium (with positive permittivity) behind which a recording device is present. This particular configuration is descriptive of the near-field optical lithography imaging system used in recent investigations~\cite{fang05,melville05}                                         

\section{Optical transfer function of the system}

\begin{figure}[t]
\centering
\centerline{\includegraphics[width=0.9\columnwidth]{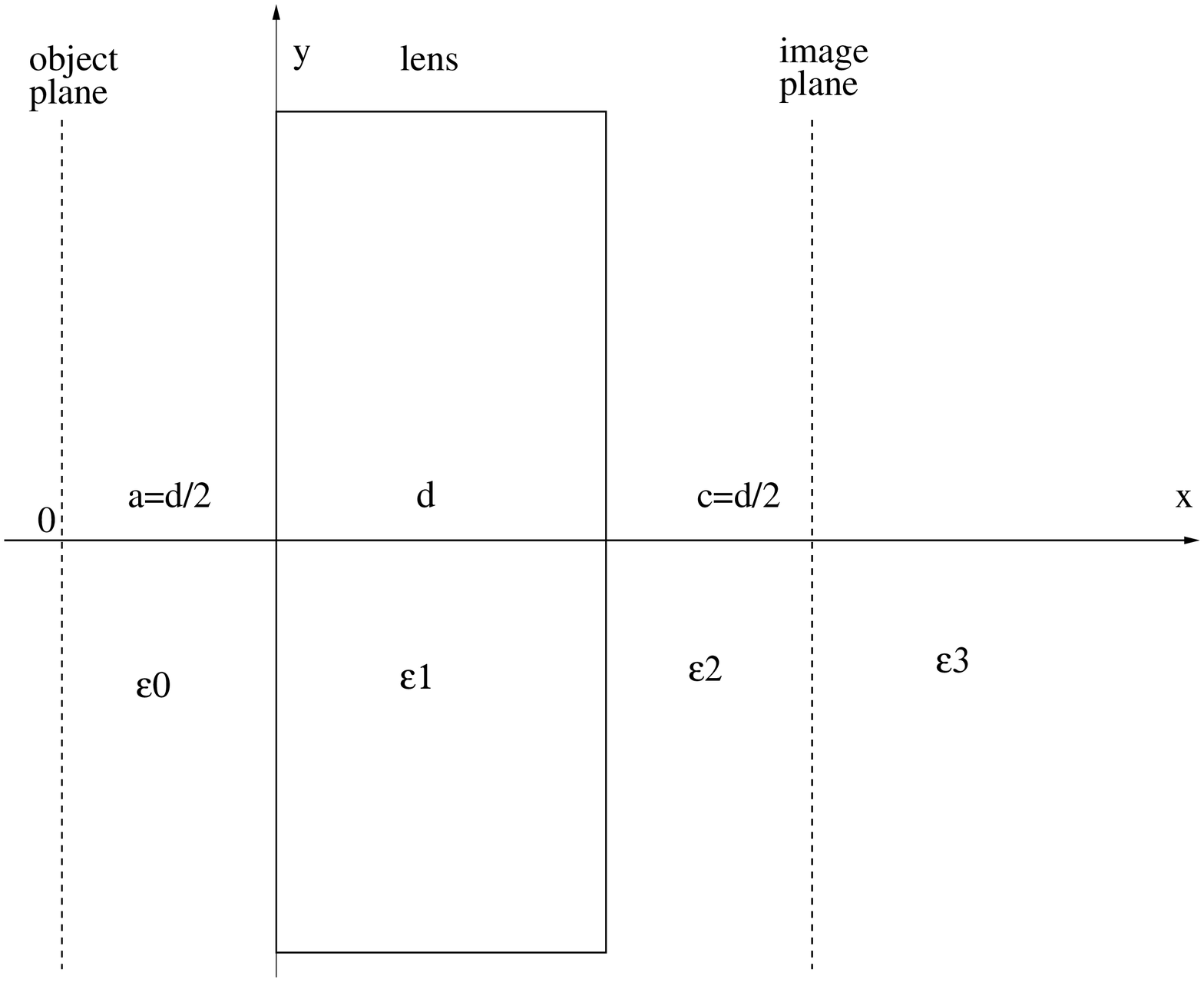}}
\caption{A schematic geometry of a system under investigation.}
\label{fig1}
\end{figure}

\begin{figure}[t]
\centering
\centerline{\includegraphics[width=0.9\columnwidth]{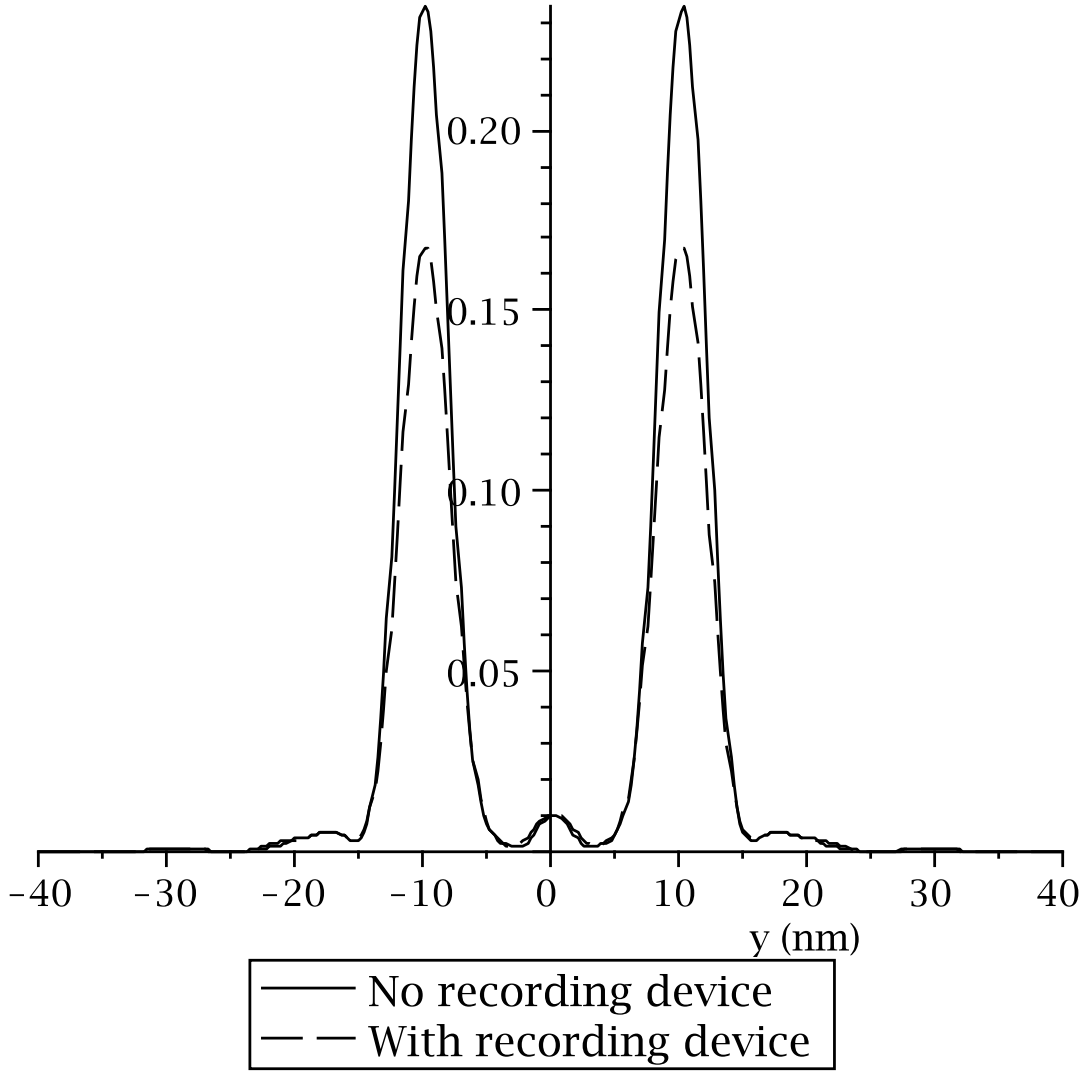}}
\caption{The image of two slits of width 2.5 nm, separated by 20 nm. The solid line represents the case when there is no recording device present (Im$\epsilon_3$=0) and the dashed line represents the situation with the detector present ($\epsilon_3$=10.47+8.0*I). The incident light wavelength is 0.52 $\mu m$, $\epsilon_0=11.7$, $\epsilon_1$=-10.47+0.2811*I, $\epsilon_2=10.47$. The thickness of the lens is $k_0d=0.07$ where $k_0$=1.2079$\times$ $10^{7}$ $m^{-1}$.}
\label{fig2}
\end{figure}

The imaging problem can be described in terms of the optical transfer function $\tau (x,k_y,\omega)$ ($k_y$ designates the in-plane wave vector directed along the surface of the material), defined as the ratio of Fourier
components of image field to object field, $B_{img}^{k_y}(x)/B_{obj}^{k_y}(0) $ (for $-\infty <k_y<\infty $) at a given imaging plane $x$. The transfer function $\tau $ can be used to find the reconstructed field in
the image plane in the form
\begin{equation}
B_{img}(x,y,t)=\int B_0(k_y)\tau (x,k_y,\omega)e^{i(k_yy-\omega t)}dk_y,
\label{eq1}
\end{equation}
where $B_0(k_y)$ is the wave vector spectrum of the source (imaged object). Thus, the ability of the system to image the object is completely determined by the optical transfer function, which in itself depends on many physical parameters of the system. In an ideal case, the transfer function should transfer all spatial harmonics equally or $\tau (x,k_y,k_0)=const.$ for $-\infty <k_y<\infty $. In reality however, the transfer function is a non-monotonous function of the wave vector $k_y$, medium material type, its thickness $d$, and the position $x$ of the imaging plane relative to the position of the object plane. It can be found by considering a $p$-polarized wave (electric vector in the plane of incidence) incident on a thin slab of thickness $d$, dielectric permittivity $\epsilon_1 <0$ and magnetic permeability $\mu_1 =1$ as shown in Figure~\ref{fig1}. The lens is separated from the source plane by the medium with permittivity $\epsilon_0$ and permeability $\mu_0=1$ and the thickness $a$. The detector is modeled as a medium with permittivity $\epsilon_3$ and permeability $\mu_3=1$. It is located at a distance $c$ from the lens. The space between the detector and the lens is filled with medium having permittivity $\epsilon_2$ and permeability $\mu_2=1$. The optical properties of this imaging system are obtained by taking the ratio of the field in the region $x>a+d+c$ to that at the object plane (in current calculations the object plane is assumed to be at $x=0$). The electromagnetic fields in each region of interest are found from solving the well known wave equation,
\begin{equation}
\epsilon \frac d{dx}\left( \frac 1\epsilon \frac{dB_z}{dx}\right) +\frac{%
\omega ^2}{c^2}\left( \epsilon -\frac{k_y^2c^2}{\omega ^2}\right) B_z=0,
\label{eq2}
\end{equation}
with a general solution having the following form,
\begin{equation}
B_z=\left( A_1e^{ikx}+A_2e^{-ikx}\right) e^{i(k_yy-\omega t)}  \label{eq2a}
\end{equation}
where $k=\omega /c\sqrt{(\epsilon -k_y^2c^2/\omega ^2)}$. The electromagnetic fields in the first medium ($x<a$) represent a sum of incident (emitted by the object at $x=0$) and reflected waves. The field in the detector ($x>a+d+c$) is modeled as a transmitted wave only. Matching solutions at different boundaries by requiring the continuity of $B_z$ and $1/\epsilon dB_z/dx$ across interfaces, one arrives at the expression for the transfer function at $x=a+d+c$ (imaging plane),
\begin{equation}
\tau (a+d+c,k_y,\omega)=\frac{8 e^{i (k_3 c+k_1 d+k_0 a)}\xi_0 \xi_1 \xi_3}{-e^{2 i (k_1 d+k_3 c)}
\Xi_{0}+e^{2 i k_3 c}\Xi_{1}+e^{2 i k_1 d}\Xi_2+\Xi_3} \label{eq3}
\end{equation}
where $\Xi_0=(\xi_0-\xi_1) (\xi_2-\xi_3) (\xi_1+\xi_3)$, $\Xi_1=(\xi_0+\xi_1)(\xi_1-\xi_3) (-\xi_2+\xi_3)$, $\Xi_2=(\xi_0-\xi_1) (\xi_1-\xi_3)(\xi_2+\xi_3)$, $\Xi_3=(\xi_0+\xi_1) (\xi_1+\xi_3) (\xi_2+\xi_3)$ and $\xi_i=k_i/\epsilon_i$. The zeros of the denominator in Eq.~\ref{eq3} define dispersion relation for the eigenmodes supported by the given system, which in general are coupled plasma surface waves running on either side of the lens. Their presence is detrimental for imaging purposes, since they disproportionately enhance the resonant spatial frequencies in the image. Presence of dissipation in the lens may dramatically reduce and widen the resonances, leading in certain cases to some improvement in the image quality\cite{fourkal07,ramakrishna02}. As shown in Ref.~\cite{ramakrishna02}, a favorable distribution of material in the asymmetric superlens system should be such that $Re\epsilon_2=Re\epsilon_3=-Re\epsilon_1$. This choice of permittivity distribution leads to weaker dependence of the transfer function on the wave vector as well as reduces the amount of the reflected wave generated between an object and the lens (the presence of large reflected wave leads to distortion of the object field, thus producing artifacts in the image). Figure~(\ref{fig2}) shows the reconstructed image of two slits of width 2.5 $nm$ each, separated by a distance of 20 $nm$, for two different cases, with and without the recording device. It should be noted here that the absence of the recording device corresponds to the case when Im$\epsilon_3$=0 (Re[$\epsilon_2$]=Re[$\epsilon_3$]).  As one can see, the presence of a dissipative recording device somewhat deteriorates the quality of an image as compared to the ideal case of no detector present. It is also worth noting that the quality of the resultant image is not only dependent on the optical properties of the imaging system but is also linked to the imaging object itself as well as its spatial spectral characteristics or information content. As mentioned earlier, the ideal imaging system should transmit all spatial components equally. In reality however, for all practically accessible lens parameters (its width, amount of dissipation, and the choice of material surrounding the lens), the optical transfer function is usually a non-monotonous function of the wave vector $k_y$ with zero asymptotic value at large $k_y$. Therefore, the quality of the reconstructed image depends on the relation between the characteristic frequency $k_{eff}^{ob}$ of the imaged object (maximum value of the wave vector $k_y$ for which the object's Fourier spectrum is not negligible) and the characteristic frequency $k_{eff}^{L}$ of the lens (maximum value of the wave vector $k_y$ for which the optical transfer function is not negligible) as well as the functional shape of the optical transfer function. The condition  $k_{eff}^{ob} < k_{eff}^{L}$ along with the optical transfer function having weak dependence on $k_y$ presents a favorable case in terms of the quality of the resultant image. As mentioned earlier, the presence of the recording device reduces the value of $k_{eff}^{L}$\cite{vinogradov05}, but if the above condition is still satisfied, the image deterioration due to the presence of the recording device will be minimal.

\section{Information content of the imaged object}
\begin{figure}[t]
\centering
\centerline{\includegraphics[angle=-90,width=0.9\columnwidth]{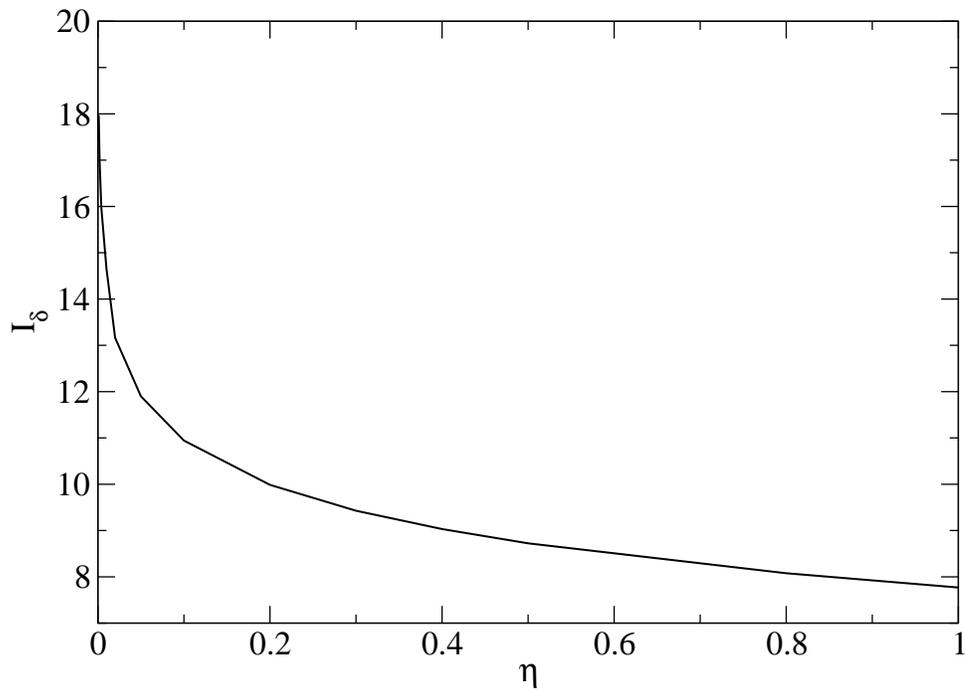}}
\caption{The information contained in the object (two slits of width 2.5 nm, separated by the distance 20 nm) versus the wave vector's sampling size $\eta=\delta/k_0$, where $k_0$=1.2079$\times$ $10^{7}$ $m^{-1}$.}
\label{fig3}
\end{figure}

In order to quantify the "complexity" of an object or its information content, we'll use the ideas from the communication theory\cite{shannon48}, specifically the definition of information through entropy,
\begin{equation}
I=-\sum_ip_i\log_2 p_i, 
\label{eq4}
\end{equation}     
where $p_i$ are the {\it a priori} probabilities of occurrence of various states related in our case to the imaging object. It is natural to define it as the probability of occurrence of $k_y^i$ spatial frequency in the Fourier spectrum of the object. In this respect, the object can be viewed as an information source which contains a message in a form of a sequence of different components of the wave vector $k_y$. This probability can be expressed in a discrete form as,
\begin{equation}
p_i=p(k_y^i,\delta)=\int_{k_y^i-\delta/2}^{k_y^i+\delta/2}P(k_y) dk_y
\label{eq5}
\end{equation}  
where $\delta$ is the sampling size and $P$ is the density probability function, which is related to object's spectral density distribution,
\begin{equation}
P(k_y)=A\left|\int_{-\infty}^{\infty}B_0(y)e^{ik_y y}dy\right|^2
\label{eq6}
\end{equation}           
where A is a normalization factor found from the condition $\int_{-\infty}^{\infty}P(k_y)dk_y=1$. Substituting Eq.~\ref{eq6} into Eq.~\ref{eq5} and subsequently Eq.~\ref{eq4}, one obtains an expression for information measure $I(\delta)$ of the imaged object (in bits) for the given degree of discretization $\delta$. Taking subsequently the limit $\delta \rightarrow 0$ provides the total information contained in the imaged object (the existence of such a limit follows from the normalization condition on the density probability function). If one is able to transfer all of this information from the object plane to the image plane and imprint it into the recording device, a perfect image will be formed. However, since the recording device in its design is a discrete system (degree of discretization is determined by the pixel size), the recorded image will never contain all of the initial information "emanating" from the object even in the presence of an ideal lens. Figure~\ref{fig3} shows the dependence of the information contained in the object (two slits of width 50 nm separated by the distance 100 nm) as a function of the sampling size $\delta$ (the wave vector's sampling size $\delta$ is related to the recording device's pixel size $\alpha$ through the following relation $\delta=2\pi \frac{\alpha}{\lambda^2}$, where $\lambda$ is the characteristic wavelength of the object's spatial inhomogeneity; it is the distance between the two slits in the example presented above). As $\delta$ increases, the information imprinted in the device decreases, reducing the image quality of the original object. This information reduction effect is due to a coarse-graining or "zooming out" procedure irrevocably introduced by observing the "world" at a finite level of resolution. The act of recording will raise the entropy of the recording device by at least the amount equal to that of the recorded information. The entropy of the observed object however should decrease by the same amount as a result of the act of observation\cite{tegmark00}. This entropy reduction of an object can be also understood from considering the relation between the entropy and information: the thermodynamic entropy is an estimate of the amount of further information needed to define the detailed microscopic state of the system. Once the observer has recorded a part of the total information content of an object, the remaining not yet observed or hidden information determines the object's entropy\cite{lloyd1989}.                                 

In the arguments presented above, we didn't account for the influence of the superlens on the object information. Since the lens in general, does not equally transfer all spatial harmonics $k_y$ but rather enhances one part of the original spectrum and suppresses the other (except for an ideal case of extremely thin lens of thickness less than 1 nm, in which case the transfer function is indeed constant over very large wave numbers), the object's spectral distribution "emanating" behind the lens is somewhat reshuffled (changing the probabilities of occurrence of $k_y^i$ spatial components), distorting the shape of the original object and changing its information content (Eq.~\ref{eq4}). The object's information reduction-distortion effect by the lens on one hand is due to already mentioned presence of surface wave resonances. On the other hand it is also due to the so-called "information hiding" effect, in which a part of the original object's information imprinted in the evanescent waves becomes hidden to the outside world as an "observer" moves away from the object's plane. This information loss is reversible in its nature, since the hidden information can be recovered by simply putting the recording device closer to the object. The presence of the lens simply allows recovering of some of the hidden information at a distance from the object (not possible in the absence of the lens). 

In addition, when there is dissipation in a lens, part of the object's information gets irreversibly lost to the observer due to energy absorption. This information does not disappear to the outside world however, but is carried away by the thermal photons radiated away from the lens.

\section{Energy and information transport in superlensing}

It would be instructive to first revisit the question of what the act of information recording is. During this process, the incoming information is imprinted in the recording device via some process of energy deposition. Thus the recording device registers the energy flux, or the normal component of the Poynting vector but not the value of the field. Therefore, the optical transfer function that has been calculated as a ratio of the field value at the imaging plane to that at the object plane (Eq.~\ref{eq3}) and explored in many previous investigations serve more of an instructive role on the nature of the evanescent field amplification rather than describing the real experimental situation. Therefore, the experimentally relevant optical transfer function should be defined as that proportional to the normal component of the Poynting vector at the imaging plane. Using the equations from the previous section, the normal component of the time averaged Poynting vector $P_x=\frac 1 2 Re[E_yB^*_z]$ is given by the following expression,
\begin{equation}
P_x=\frac 1 2 Re[\int \int \xi_3(k'_y)\tau(k'_y)B_0(k'_y)\tau^*(k'_y-k_y)B^*_0(k'_y-k_y)e^{-ik_yy}dk'_ydk_y]
\label{eq7}
\end{equation} 
For the case when the incident wave is a plane wave, so that $B_0(k_y)=B_0\delta(k_y-k^0_y)$, the $x$ component of the Poynting vector simplifies to $P_x=\frac 1 2 Re[\xi_3]\left|\tau\right|^2\left|B_0\right|^2$. Thus, the optical transfer function in this case may be defined as,
\begin{equation}
\Upsilon(a+d+c,k^0_y,\omega)=\left[\left|\tau\right|^2Re[\xi_3]\right]_{k_y=k^0_y}.
\label{eq8}
\end{equation} 
It is interesting to note that the energy flux optical transfer function $\Upsilon$ is equal to zero when $Re[\xi_3]=0$, which for the evanescent part of the spectrum occurs when $Im[\epsilon_3]=0$ (lossless recording device). For the propagating part of the spectrum ($k_y<\omega/cRe[\epsilon_3]$) however, $\Upsilon\neq 0$ for any value of the recording device's permittivity $\epsilon_3$, confirming the point made in the introduction section that presence of the recording device does not alter the object's information content transported by the propagating part of the photon spectrum. Therefore, the dissipation in the recording device plays a fundamental role in transferring energy as well as fine-structure information about the object (carried by the evanescent components of the spectrum). If there is no recording device present, there is no energy or information flow in the system for all of the modes with $k_y>\omega/cRe[\epsilon_3]$. As soon as the recording device is introduced, the flow of energy and information (for modes with $k_y>\omega/cRe[\epsilon_3]$) from the object to IRD starts. Therefore, the dissipation in the recording device in a way materializes the fine structure information about an object.     

\begin{figure}[t]
\centering
\centerline{\includegraphics[angle=-90,width=0.9\columnwidth]{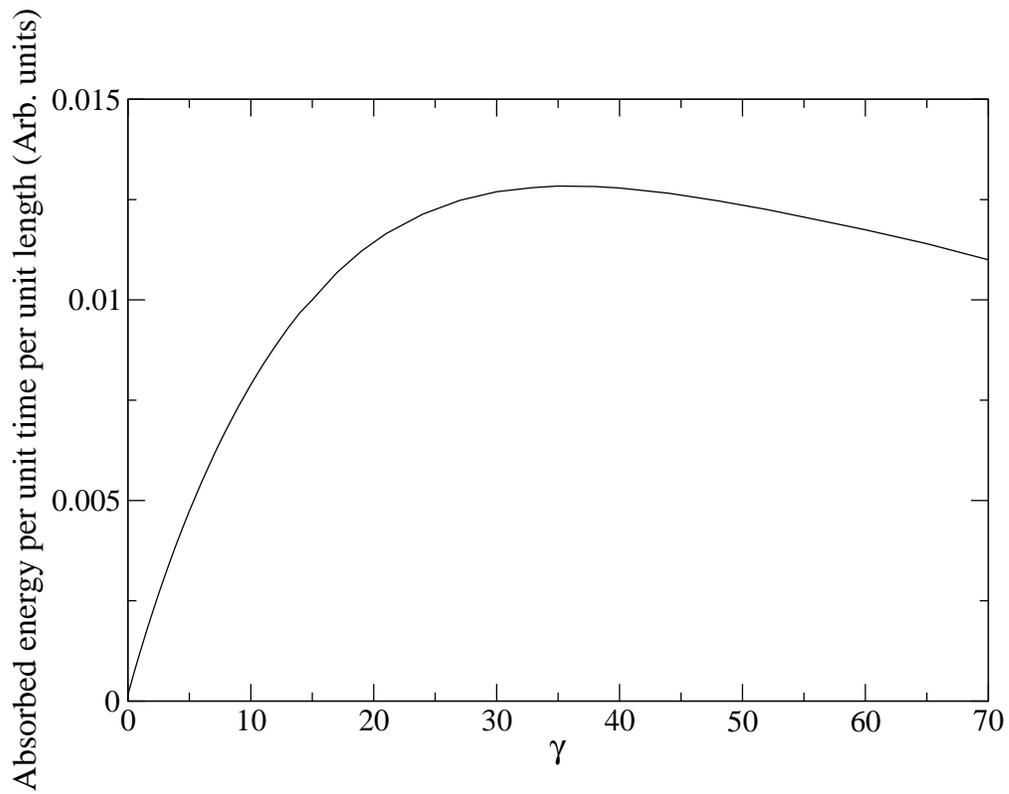}}
\caption{The energy absorbed in the detector (per unit length and time) versus its dissipation constant  ($\gamma=Im\epsilon_3$) for the case of two slits imaged using light with 0.52 $\mu m$ wavelength. The parameters of the imaging system are the same as those given in Fig. 2 }
\label{fig4}
\end{figure}

\begin{figure}[t]
\centering
\centerline{\includegraphics[angle=-90,width=0.9\columnwidth]{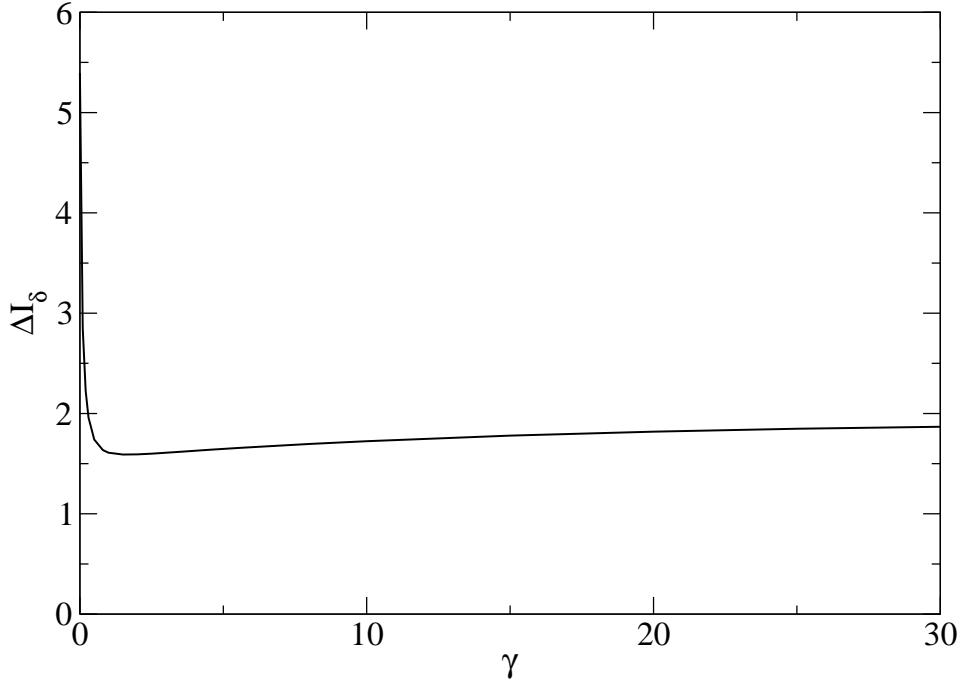}}
\caption{The information distortion function $\Delta I_{\delta}$=($I_{\delta}-I_{\delta}^{img}$) versus the dissipation constant $\gamma$ of the recording device for the case when two slits are imaged using light with 0.52 $\mu m$ wavelength. The parameters of the imaging system are the same as those given in Fig. 2. The degree of discretization $\delta=0.01*k_0$, where $k_0=1.2079\times 10^{7}$ $m^{-1}$.}
\label{fig5}
\end{figure}

Moreover, not only does the Eq.~\ref{eq7} determine the amount of energy transferred, but it also establishes the amount of information "incident" on the recording device. The total energy absorbed in the device can be found by integrating expression \ref{eq7} over the whole detector's length in $y$ direction to give,
\begin{equation}
\frac{E}{t L}=\frac 1 2 \int \left|\tau(k_y)\right|^2Re[\xi_3(k_y)]\left|B_0(k_y)\right|^2dk_y,
\label{eq9}
\end{equation}            
where $E/L$ is the total energy absorbed per unit detector's length and $t$ is the recording time. It would be useful to define the energy spectral density function $S_x=\frac 1 2\left|\tau(k_y)\right|^2Re[\xi_3(k_y)]\left|B_0(k_y)\right|^2$, which may be further used to define the density probability function for occurrence of $k_y$ spatial frequency at the imaging plane,
\begin{equation}
P^{img}(k_y)=\Psi \frac 1 2 \left|\tau\right|^2Re[\xi_3]\left|B_0(k_y)\right|^2= \Psi \cdotp S_x(k_y),     
\label{eq10}           
\end{equation}
where $\Psi$ is the normalization constant. Using the above equation as well as the definition of information~(\ref{eq4}), one can characterize the amount of object's information $I^{img}_{\delta}$ at the imaging plane,
\begin{equation}
I^{img}_{\delta}=-\sum_{k_y^i} p^{img}_i\log_2[p^{img}_i]
\end{equation}  
where $p^{img}_i=\frac 1 2\Psi \int_{k_y^i-\delta/2}^{k_y^i+\delta/2}\left|\tau\right|^2Re[\xi_3]\left|B_0(k_y)\right|^2 dk_y$.
Next, one may define the function $\Delta I_{\delta}=\left(I_{\delta}-I^{img}_{\delta}\right)$ (let us call it information distortion function) that quantifies the difference between the "true" information content of an object and the transferred information content. Clearly, the goal of the imaging system would be to provide maximum energy flow between the object and imaging planes, minimizing the information distortion function $\Delta I_{\delta}$. In reality however, the "true" information content of an object $I_{\delta}$ is unknown to the observer (otherwise, what is the point of imaging the object) and minimization procedure is not possible. In this situation the only viable approach is to find a set of parameters for which the optical transfer function  $\Upsilon$ of the imaging system stays constant over wider range of the wave numbers $k_y$. Nonetheless, it would be instructive to see how variation of a single variable (dissipation in the recording device) in the imaging system influences both, the total energy absorbed in the detector and the information distortion function $\Delta I_{\delta}$ for the case when the initial object is a two-slit system. Figures~\ref{fig4} and \ref{fig5} show both quantities versus the imaginary part of the recording device's permittivity $\gamma$ ($\epsilon_3=Re[\epsilon_3]+i*\gamma$). As one can see, there is a non-monotonous dependence of both quantities on the amount of dissipation in the recording device. Furthermore, there also seems to be a range of values for the dissipation constant where the energy flow and information distortion function reach their corresponding maximum and minimum values. This indicates the existence of some advantageous device operation regimes, warranting future investigations. In addition, figure 6 shows the $x$ component of time-averaged Poynting vector as a function of $y$ coordinate, (calculated at the position of the recording device) for the case of two slits imaged by light with wavelength 0.52 $\mu m$. As one can see, the actual image "imprinted" in the device is somewhat different from that shown in Figure~\ref{fig2}, that was calculated using only the transfer function for the magnetic field.                           

Using the universal bound on the energy cost per bit of information transferred, one arrives at the following inequality for the recording time,
\begin{equation}
t^2>\Psi \frac{\hbar \ln[2]}{\pi L}I^{img}_{\delta}
\end{equation}  
Since both, the normalization constant $\Psi$ and the amount of information available at the imaging plane 
$I^{img}_{\delta}$ are functions of the recording device's dissipation constant, the above inequality can in general be written as the relation between the recording time and the dissipation constant of the recording device (the inequality would not be universal but object specific). 

\section{Summary}

In conclusion, we have shown that the recording device plays a fundamental role in the image formation and registration. Not only does it change the optical transfer function of the imaging system, but it actually determines what the observer will see of the imaged object, depending on the physical characteristics of the recorder itself. In the case studied here, the information contained in the observed object depends on the imaging system's characteristics, and the question of how much the recording device affects the result of the observation becomes important. The critical influence of the recording device comes from the requirement of the presence of both (incident and reflected) evanescent waves to provide a finite energy flux between the object and the recording device. As a result, the issue of the properties of
the recording device becomes non-trivial for the superlensing where a significant information is transmitted in the evanescent part of the spectrum. On a more philosophical note, we could not help noticing that the results of this study have certain parallels with the "measurement problem" in quantum mechanics where the observed property of an object emerges from the actual act of measurement.   
       
\section*{Acknowledgment} 
This work is in part supported by Strawbridge Family Foundation, NSERC
Canada and  AFOSR Award \#FA9550-07-1-0415.

\begin{figure}[t]
\centering
\centerline{\includegraphics[width=0.9\columnwidth]{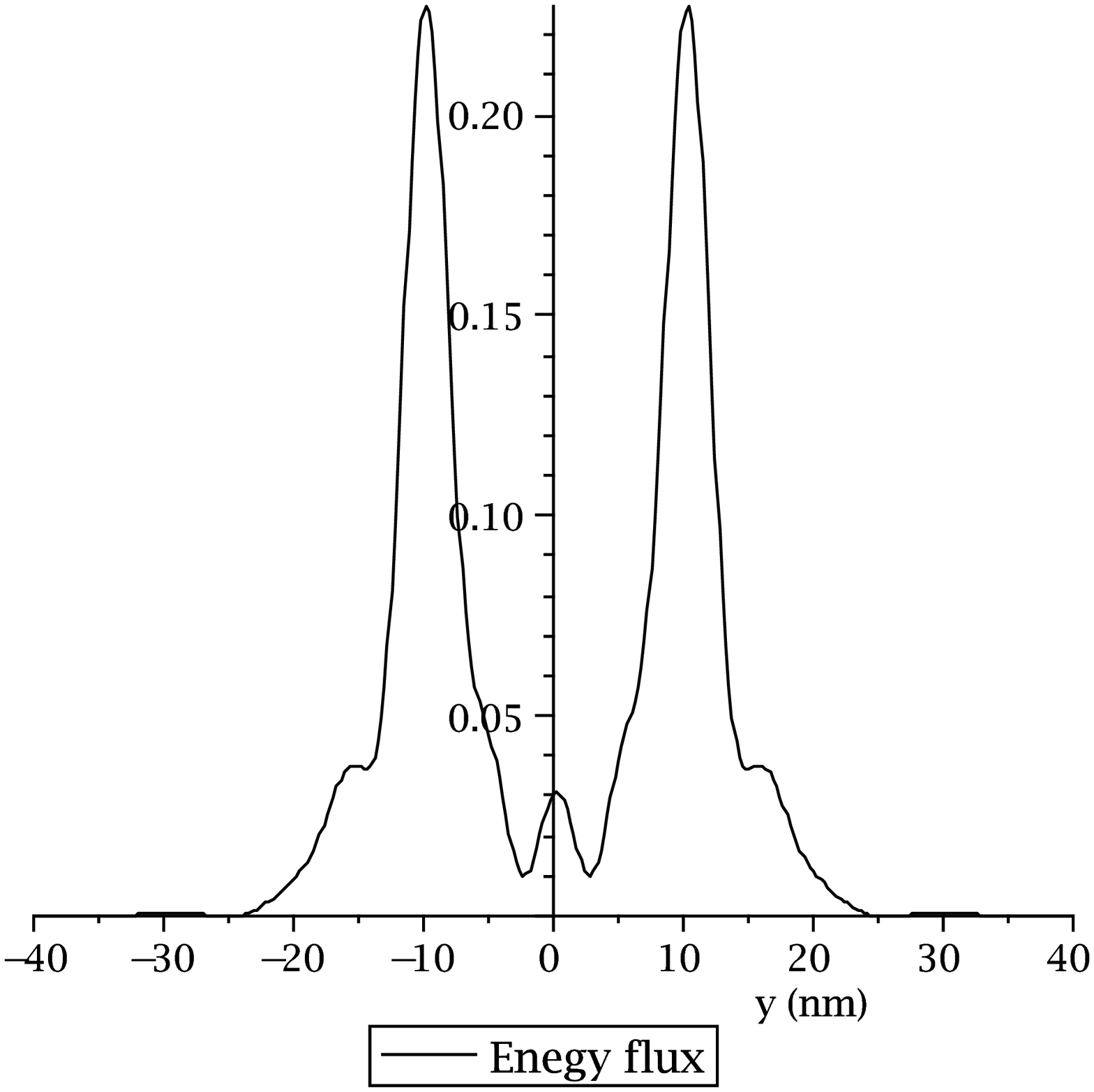}}
\caption{The calculated time-averaged energy flux (the $x$ component of the Poynting vector) of an image of two slits of width 2.5 nm, separated by the distance 20 nm. The parameters of the imaging system are the same as those given in Fig. 2}
\label{fig6}
\end{figure}


\begin{thebibliography}{10}

\bibitem{kindel75}
J.~M. Kindel, K.~Lee, and E.~L. Lindman.
\newblock Surface wave absorption.
\newblock {\em Phys. Rev. Lett.}, 34:134--138, 1975.

\bibitem{dragila85}
R.~Dragila, B.~Luther-Davies, and S.~Vukovic.
\newblock High transparency of classically opaque mettalic films.
\newblock {\em Phys. Rev. Lett.}, 55:1117--1120, 1985.

\bibitem{bliokh05}
Y.~Bliokh, J.~Felsteiner, and Y.~Slutsker.
\newblock Total absorption of electromagnetic wave by an overdense plasma.
\newblock {\em Phys. Rev. Lett.}, 95:165003--1--165003--4, 2005.

\bibitem{fang05}
N.~Fang, Hyesog Lee, Cheng Sun, and Xiang Zhang.
\newblock Sub-diffraction-limited optical imaging with a silver superlens.
\newblock {\em Science}, 308:534, 2005.

\bibitem{barnes03}
W.~Barnes, A.~Dereaux, and T.~Ebbesen.
\newblock Surface plasmon subwavelength optics.
\newblock {\em Nature}, 424:824--830, 2003.

\bibitem{maier05}
S.~Maier.
\newblock Plasmonics - towards subwavelength optical devices.
\newblock {\em Current Nanoscience}, 1:17--23, 2005.

\bibitem{smith04}
D.~Smith, J.~Pendry, and M.~Wiltshire.
\newblock Metamaterials and negative refractive index.
\newblock {\em Science}, 305:788--792, 2004.

\bibitem{pendry00}
J.~B. Pendry.
\newblock Negative refraction makes a perfect lens.
\newblock {\em Phys. Rev. Lett.}, 85:3966--3969, 2000.

\bibitem{gomez-santos03}
G.~Gomez-Santos.
\newblock Universal Features of the time evolution of evanescent modes in a left-handed perfect lens.
\newblock {\em Phys. Rev. Lett.}, 90:077401, 2003.


\bibitem{bekenstein81}
J. D.~Bekenstein.
\newblock Energy cost of information transfer.
\newblock {\em Phys. Rev. Lett.}, 46:623--626, 1981.

\bibitem{bekenstein88}
J. D.~Bekenstein.
\newblock Communication and energy.
\newblock {\em Phys. Rev. A}, 37:3437--3449, 1988.

\bibitem{shannon48}
C.~Shannon.
\newblock A mathematical theory of communication.
\newblock {\em Bell System Technical Journal}, 27:623--656, 1948.

\bibitem{kolokolov92}
A.~Kolokolov and G.~Strotsky.
\newblock Interference of the evanescent components of the electromagnetic field.
\newblock {\em Phys. Uspekhi}, 162:165, 1992.

\bibitem{fourkal07}
E.~Fourkal, I.~Veltchev, C.~Ma, and A.~Smolyakov.
\newblock Resonant transparency of materials with negative permittivity.
\newblock {\em Physics Letters A}, 361:277--282, 2007.

\bibitem{podolskiy05}
V.~A. Podolskiy, N.~A. Kuhta, and G.~W. Milton.
\newblock Optimizing the superlens: Manipulating geometry to enhance the
  resolution.
\newblock {\em Applied Physics Letters}, 87(23):3, 2005.
\newblock Times Cited: 14.

\bibitem{vinogradov05}
A.~Vinogradov and A.~Dorofeenko.
\newblock Destruction of the image of the pendry lens during detection.
\newblock {\em Optics Communications}, 256:333--336, 2005.

\bibitem{melville05}
D.~Melville and R.~Blaikie.
\newblock Super-resolution imaging through a planar silver layer.
\newblock {\em Opt. Exp.}, 13:2127--2134, 2005.

\bibitem{ramakrishna02}
S.~Ramakrishna, J.~Pendry, D.~Schurig, D.~Smith, and S.~Schultz.
\newblock The assymetric lossy near-perfect lens.
\newblock {\em Journal of Modern Optics}, 49:1747--1762, 2002.

\bibitem{tegmark00}
M.~Tegmark.
\newblock Importance of quantum decoherence in brain processes.
\newblock {\em Phys. Rev. E}, 61:4194--4206, 2000.

\bibitem{lloyd1989}
S.~Lloyd.
\newblock Use of mutual information to decrease entropy: Implications for the
  second law of thermodynamics.
\newblock {\em Phys. Rev. A}, 56:5378--5386, 1989.



\end{thebibliography}
\end{document}